# XML Multidimensional Modelling and Querying

Serge Boucher, Boris Verhaegen, Esteban Zimányi

Department of Computer & Decision Engineering (CoDE)
Université Libre de Bruxelles (U.L.B.)
{sboucher,boverhae,ezimanyi}@ulb.ac.be

**Abstract.** As XML becomes ubiquitous and XML storage and processing becomes more efficient, the range of use cases for these technologies widens daily. One promising area is the integration of XML and data warehouses, where an XML-native database stores multidimensional data and processes OLAP queries written in the XQuery interrogation language. This paper explores issues arising in the implementation of such a data warehouse. We first compare approaches for multidimensional data modelling in XML, then describe how typical OLAP queries on these models can be expressed in XQuery. We then show how, regardless of the model, the grouping features of XQuery 1.1 improve performance and readability of these queries. Finally, we evaluate the performance of query evaluation in each modelling choice using the eXist database, which we extended with a grouping clause implementation.

## 1 Introduction

Among all the assets that an organisation possesses, the data it has gathered over the years is fast becoming the most valuable. Helping to make the most of this value is data warehousing technology, which stores and organises operational data in a way tailored to facilitate analysis and decision making. Typically, this involves a multidimensional model: the dataset is seen as a collection of *facts*, or atomic pieces of information (e.g., orders for products), containing *measures* (e.g., quantity or price) that can be analyzed along different *dimensions*, or perspectives (e.g., customer localisation or product family). Each dimension may contain *hierarchies*, composed of several levels, allowing the measures to be analyzed at different granularities (e.g., the localisation dimension can have levels such as city, country, and continent). Such a multidimensional model lends itself to numerous types of query that have significant business value, and their processing is called either multidimensional analysis or Online Analytical Processing (OLAP) [8].

Simultaneously with the growth in data storage and collection which fuelled innovation in data warehousing, there has been a huge increase in data interchange, which lead to the creation of standards, notably the XML eXtended Markup Language. Huge quantities of data are now available in XML format and tools to process it are improving daily. Most notably, XML native databases, especially designed to store and query XML data efficiently, and the XQuery interrogation language [7], which is the XML equivalent of SQL, are now mature enough to be used in a production environment.

The usual approach for warehousing XML data is to go through a conversion phase and store it in relational databases. However, as more and more information starts out in XML form, an interesting prospect is the implementation of a data warehouse using only XML technologies: XML documents model facts and dimension hierarchies, these documents are stored in an XML native database, and XQuery is used for analysis of this multidimensional data.

The main contribution of this paper is a discussion of issues arising during implementation of a data warehouse using XML technologies: we present different approaches for multidimensional modelling in XML, discuss the implementation of common queries on those models using XQuery, and evaluate the performance implications of each modelling choice on the open-source XML native database eXist [14].

The remainder of this paper is organised as follows: Section 2 shows various ways to model multidimensional data in XML. Section 3 presents how typical OLAP queries on these XML data warehouses can be answered using XQuery. Section 4 contains a performance analysis and a comparison of those models for various multidimensional queries. Finally, Section 5 concludes the paper and points to further research work.

## 2    Multidimensional Modelling with XML

In this section, we present three broad approaches for modelling multidimensional data in XML documents. The first one, referred to as *flat model*, puts all information about a fact and its dimension hierarchies in the same XML element. The second, referred to as *hierarchical model*, organises facts and dimensions in distinct XML elements and use XML parent-child relationships to model dimension hierarchies. The last one, the *XCube model* [10], use identifiers to model the imbrication of levels in dimension hierarchies.

### 2.1    Running Example

All along this paper, concepts will be illustrated with the same example data warehouse, containing a set of facts and two dimension hierarchies. Fig. 1 presents our running example using the MultiDim formalism presented in [13]. In this example, facts are orders of products whose measures are *quantity* and *price*. Dimensions are the *product* and the *customer* of an order. Those dimensions are organised in hierarchies, respectively *customer-country-continent* and *product-category-family*. Those hierarchies are *strict* [12], since an instance of a given level is related to exactly one instance of the upper level: for example, a country is related to only one continent.

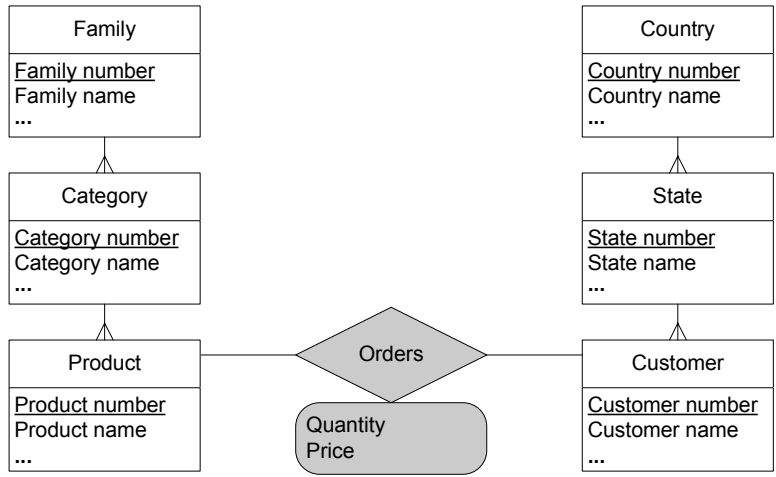

**Fig. 1.** MultiDim schema of our running example

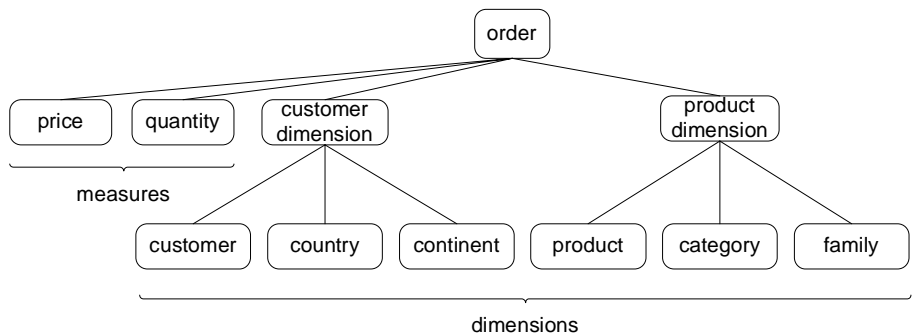

**Fig. 2.** Schema of a fact modelled with flat model

## 2.2 Flat Model

The flat model modelling strategy represents each fact under a single XML element, with measures and dimension hierarchies as subelements of this fact. Dimension hierarchies are expressed in a flat way similar to a denormalised star schema in a relational model, as is often used for physical storage of data warehouses in relational databases. This is the model used among others in [1]. Below is an example of a fact represented in a flat model and Fig. 2 shows a graphical representation of the corresponding XML hierarchy.

```
<order>
  <price>125,67</price>
  <quantity>3</quantity>
  <customer_dimension>
    <customer>Jim</customer>
```

```
    <country>BE</country>
    <continent>EU</continent>
  </customer_dimension>
  <product_dimension>
    <product>Table</product>
    <category>Kitchen</category>
    <family>Furnitures</family>
  </product_dimension>
</order>
```

Thus, all information about a fact and its dimensions are stored in the same XML element. As we will see in the next section, this allows straightforward and efficient querying because no join operation is needed to obtain hierarchical levels. Facts represented in a flat model are also very human readable.

On the other hand, the flat model introduces massive redundancy in dimension data. For example, the information about the category and family of a product is repeated for all orders of that particular product. Besides the inefficient use of storage space, this redundancy could cause maintenance difficulties when the allocation of products to categories and/or categories to familes evolve, making update queries on this information complicated and expensive. Another inconvenience is that the flat model does not express hierarchical level imbrications. For example, the XML fragment above does not make it clear that a product category is enclosed in a product family. While this could arguably be enforced using an external schema, it still means that important data model information that could be embedded in the data itself ends up stored elsewhere.

This semantic deficiency can be alleviated by slightly changing the model, as suggested by Boussaid *et al.* [5]. In their model, each fact still contains all dimension hierarchies, but they are represented in a tree from the lowest to the highest hierarchical level. The customer dimension hierarchy for the fact presented above would be presented as follows:

```
<product name="Table">
  <category name="Kitchen">
    <family name="Furniture"/>
  </category>
</product>
```

This representation of dimensions resolves the semantic problem found in the flat model. However, concerns about redundancy still stand.

### 2.3 Hierarchical Model

In this paper, we propose a hierarchical model in which data redundancy is avoided by storing facts and dimension hierarchies in distinct XML elements. A fact is represented by a XML element containing only its measures and references to the lowest levels of its dimension hierarchies, as shown below:

```
<order>
  <price>125,67</price>
```

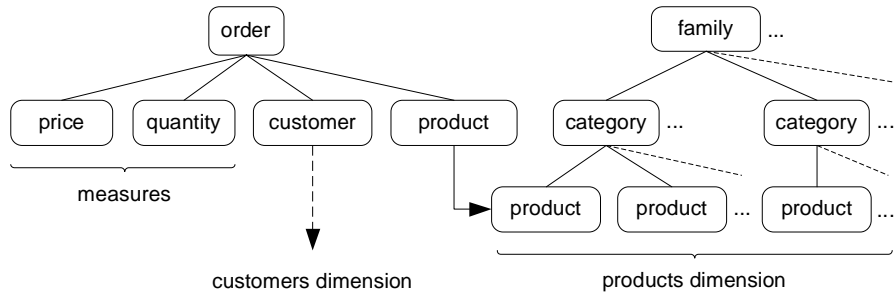

**Fig. 3.** Schema of a fact represented using the hierarchical model

```
    <quantity>3</quantity>
    <customer ref="c42"/>
    <product ref="p98"/>
</order>
```

Each dimension hierarchy is then represented by an XML element wherein imbrication between hierarchical levels is expressed with XML parent-child relations. This model thus makes elegant use of XML's tree-like structure to represent dimension hierarchies. A leaf of this tree corresponds to the lowest level of the corresponding dimension hierarchy, represented by an identifier. Fig. 3 shows a graphical representation of this model and an extract of the product hierarchy is given below:

```
<family name="Furniture">
  <category name="Kitchen">
    <product name="Table" id="p98"/>
    ...
  </category>
  ...
</family>
```

The main advantage of modelling dimensions in this way is that their use of XML's natural structure makes traversing hierarchies using XML query languages like XPath and XQuery very easy. Furthermore, hierarchical levels imbrication is clearly expressed and no redundancy is introduced in dimension data.

However, as we will see in the next sections, queries on those documents are made more complex and less efficient by the use of relational links between facts and dimensions.

### 2.4 XCube Model

XCube [10] is another formalism proposed by Hümmer *et al.* to represent multidimensional data in XML documents. In this model, each fact and each dimension level is represented by a distinct XML element. The main difference between our hierarchical model and XCube is that links between levels of a dimension hierarchy are not expressed by parent-child relationships but by identifiers. Facts (*XCubeFacts*) are represented by XML elements containing measures and links to the lowest levels of their dimension hierarchies as in our hierarchical model. Dimensions (*XCubeDimensions*) are represented by XML elements, each representing a hierarchical level. Each instance of a level is linked to the parent level instance by identifiers. Fig. 4 presents a graphical representation of this model and a fragment of the customer dimension hierarchy is shown below:

```
<level id="family">
  <node id="fam1" name="Furniture"/>
  ...
</level>
<level id="category">
  <node id="cat77" name="Kitchen">
    <rollUp toNode="fam1" level="family"/>
  </node>
  ...
</level>
<level id="product">
  <node id="p98" name="Table">
    <rollUp toNode="cat77" level="category"/>
  </node>
  ...
</level>
```

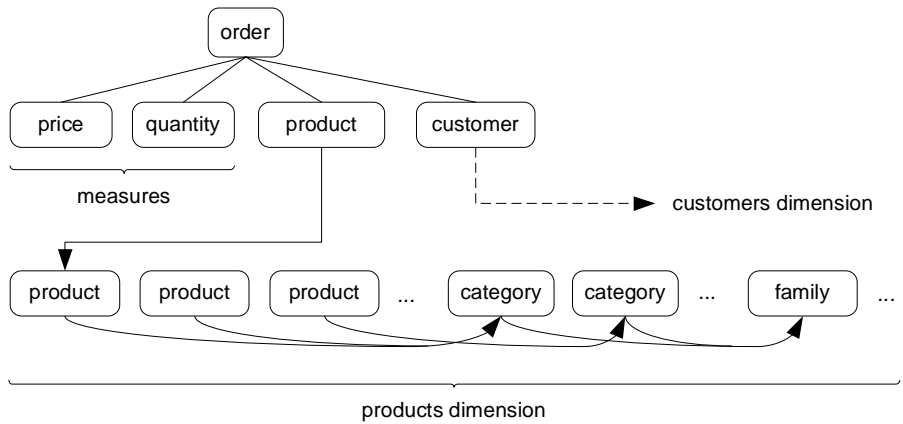

**Fig. 4.** Schema of a fact modelled with the XCube model

Like our hierarchical model, XCube does not introduce redundancy in dimension data. However, the use of identifiers between hierarchical levels instead of XML parent-child relations makes traversing dimension hierarchies more difficult. As we will show in the next sections, those links involve many joins in queries to traverse the dimension hierarchies, making queries complex to write and slow to execute.

## 3 Multidimensional Querying with XQuery

In the previous section, we have shown three ways to store a simple data warehouse in XML format. In this section, we compare querying efficiency of these three models by showing how to express typical multidimensional analysis (OLAP) queries on them using XQuery [7], a query language for XML data similar to SQL and developed by the W3C. We consider the following query for the comparison:

**Query 1.** *Give the number of ordered products by country and product category.*

### 3.1 Flat Model Querying

The translation of Query 1 in XQuery 1.0 for our running example modelled using the flat model is shown below:

```
for $category in distinct-values(//category)
  for $country in distinct-values(//country)
    let $facts := //order[(//category eq $category) and
                          (//country eq $country)]
    return
    <group>
      <category>{$category}</category>
      <country>{$country}</country>
      <sum>{sum($facts/quantity)}</sum>
    </group>
```

To answer this query using XQuery 1.0, we first have to obtain all combinations of categories and countries, using a double `for` loop which iterates over all orders and find distinct values of categories and countries. Afterward, we look for orders matching each of these combinations of country and category. This is done via an XPath expression which uses two value comparisons. Each iteration returns an XML element containing the sum of orders for the current combination of country and category.

This kind of query suffers from two problems. Firstly, they are difficult to read and write because they require an awkward search for combinations of dimension levels before the actual query on facts. Secondly, while answering this query efficiently only requires a single document traversal, a straightforward implementation traverses the whole document once for each dimension we want to analyse, plus one final time for the actual query. This is unacceptably slow for huge documents, and automatically optimising those traversals away is very difficult as shown in [1].

### 3.2 Hierarchical Model Querying

The translation of Query 1 in XQuery 1.0 for our running example modelled using our hierarchical model is shown below:

```
for $category in //category
  for $country in //country
    let $facts := /orders/order[(product/@ref  = $category//@id)
                                and (customer/@ref = $country//@id)]
    return
    <group>
      <country>{$country/@name}</country>
      <category>{$category/@name}</category>
      <sum>{sum($facts/quantity)}</sum>
    </group>
```

The general principle of this query is the same as the previous one: we have to iterate over every combination of country and category. There are, however, two important differences. Firstly, we didn't have to determine distinct values of dimension levels to find distinct combinations of them because each possible value appears in only one XML element. Secondly, XPath predicates use more complex joins because each fact contains only an identifier to the leaf level of its dimension hierarchy. In order to find orders corresponding to a category of products, we thus have to find each product of the current category and afterwards join it with facts containing only product identifiers. This join is expressed using XPath general comparison (=) which is true only if at least one element of the left operand is equal to at least one element of the right operand.

This query has similar problems to the previous one on the flat model. Firstly, it is difficult to read and write since it starts by determining all combinations of the involved dimensions and then aggregate facts, which is unnatural. Even more, XPath's general comparison is a more time-consuming operation than the simple value comparison used in the flat model because we have to compare a set (the facts) with another set (the dimension hierarchy leafs) instead of a set with a single value (the instance of the dimension hierarchy level). Those complex joins make XQuery 1.0 queries on hierarchically modelled documents significantly slower for large documents.

### 3.3 XCube Model Querying

The translation of Query 1 in XQuery 1.0 for our running example modelled using the XCube model is shown below:

```
for $category in //level[@id eq "category"]/node
  for $country in //level[@id eq "country"]/node
    let $prodId :=//level[@id eq "product"]
                   /node[rollUp/@toNode eq $category/@id]/@id
    let $cusId := /level[@id eq "customer"]
                   /node[rollUp/@toNode eq $country/@id]/@id
    let $facts := //cell[dimension[@id="products"]/@node = $prodId and
```

```
                              dimension[@id="customers"]/@node = $cusId]/fact
    return
    <group>
      <category>{$category/@name}</category>
      <country>{$country/@name}</country>
      <sum>{sum($facts/quantity)}</sum>
    </group>
```

The logic of this query remains the same as the previous ones: an iteration for each combination of category and country. The main difference is that traversing dimension hierarchies requires relational joins between hierarchy levels instead of a simple traversal of the XML tree, as is the case with the hierarchical model.

This kind of query is even more difficult to write than the previous ones and the quantity of links needed to traverse a hierarchy will lead to poorer performance than the other two models as we will see in the next section.

### 3.4  Grouping clause for XQuery

As shown above, common OLAP queries have performance and readability issues when expressed in XQuery 1.0. Time complexity of those queries increases significantly with the number of dimensions we want to analyse because they require an additional loop for each dimension. Moreover, starting the query by determining all combinations of those dimensions before aggregating the facts seems unnatural.

In the relational world, these problems are elegantly solved using the SQL `GROUP BY` [9] clause. Although the XQuery 1.0 [7] recommendation does not contain any equivalent construct, Beyer *et al.* [1] and Borkar *et al.* [3] have proposed extending XQuery with a grouping clause, and lately the W3C published a working draft about the future XQuery 1.1 [6] which will offer grouping and windowing functionalities. As we discuss in the next section, we have implemented a similar grouping clause in the XQuery engine of an XML native database.

Such a grouping clause allows writing this kind of query in a more natural way and also makes optimising them much easier. With the XQuery 1.1 syntax, our query for the flat model becomes:

```
for $fact in //order
  let $category := $fact//category
  let $country := $fact//country
  group $fact by $category, $country
  return
  <group>
    <category>{$category}</category>
    <country>{$country}</country>
    <sum>{sum($fact/quantity)}</sum>
  </group>
```

The readability advantage of this new syntax is obvious: since we do not have to compute all combinations of countries and categories beforehand, the query is more succinct and its intent is immediately clear.

As we will see in the next section, this grouping clause also improves efficiency of query answering and allows response time to remain constant with the number of dimensions we want to analyse while growing linearly with the size of the database.

## 4 Experimentation

In order to compare our models and queries, we have produced three datasets of various sizes (1K, 10K and 100K facts) for each model (flat, hierarchical and XCube). We have also added support for a grouping clause in the XQuery implementation used by eXist [14], an open source XML native database. This implementation has been integrated in the current version of eXist since release 1.2. For each model we have written eight queries (with and without grouping clause) with different numbers of results expected and with different numbers of dimensions. Datasets are stored in different collections in the eXist database and each element or attribute on which we make joins is indexed. To account for initial caching time, each query is executed four times and we use the average of the last three executions as execution time.

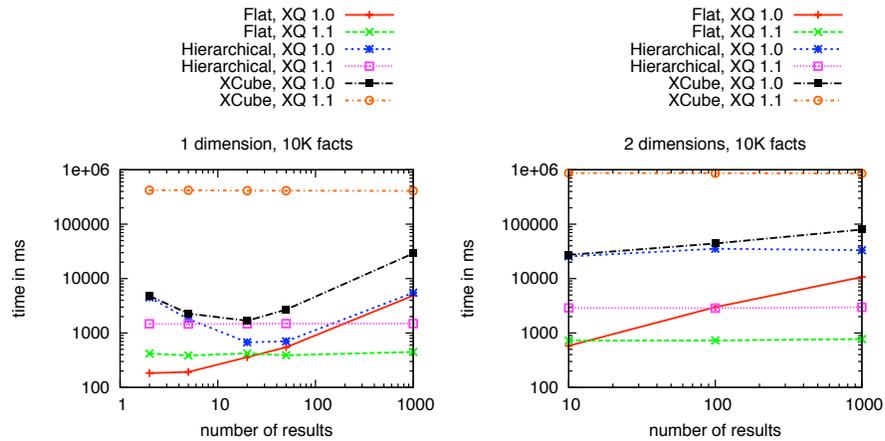

**Fig. 5.** Execution time by number of groups

Fig. 5 shows the time of execution of our queries against the number of results/groups on the facts dataset containing 10K facts for all models and for queries written with and without using the grouping clause.

A first observation we could make is that queries on the flat model are generally faster than those on hierarchical and XCube models. This is due to the computational complexity of joins needed by queries on other models because of the separation between facts and dimensions. Queries on XCube are slower than those on the hierarchical model because of the relational joins needed to traverse dimension hierarchies.

Secondly, we could also note that execution time of flat model queries without grouping clause increases linearly with the number of results/groups. Answering time for queries on the other models which do not use the grouping clause also increases with the number of groups but they present a local minimum which is due to eXist's indexing strategy: eXist's value index and XQuery general comparison implementation are optimised for joins comparing a small number of results with regards to the size of the database.

As expected, answering time for queries using the grouping clause is only dependent on the number of dimensions and the size of the database. Moreover, time seems to increase linearly with the number of dimensions and these queries are most often faster than the equivalent ones without grouping clause, except for the XCube model due to the number of joins needed in each iteration for those queries.

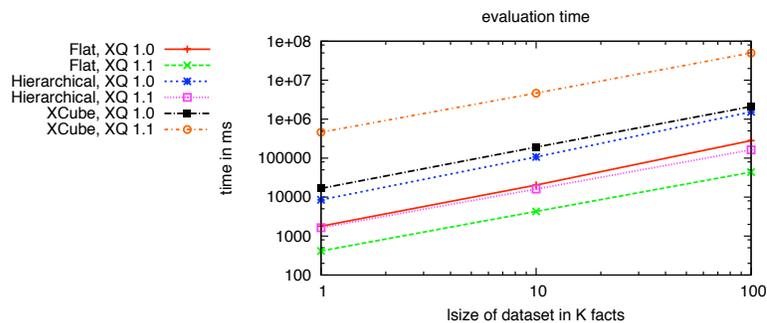

**Fig. 6.** Total execution time by size of dataset

With respect to scalability, our experimentation has shown that the execution time of all queries increases linearly with the size of the database, as shown in Fig. 6. We have also observed that the gradient of the slope of queries with grouping clause is less than those of queries in XQuery 1.0. Perhaps most importantly, the second overall fastest result was obtained using the hierarchical model with grouping clause. The implementation of this clause thus allows us to reach querying performance competitive with traditional flat models while using a richer model that decreases space requirements and improve query readability.

## 5 Related Work

This paper builds on previous work on XML modelling of multidimensional data. Bordawekar *et al.* [2] proposed a model similar to our hierarchical model, but which only permits to represent one dimension hierarchy. Beyer *et al.* [1]

proposed using a grouping clause in XQuery but only considered a flat model for their use cases and performance evaluation. Boussaid *et al.* [5] proposed a variant of the flat model that preserves the semantics of dimension hierarchies, without addressing the redundancy problems of this model.

On the specific problem of querying multidimensional XML data, Bordawekar *et al.* identified in [2] important differences between relational and XML data regarding data analysis. They also introduced grouping, cube and roll-up operators adapted for XML data. Paparizos *et al.* studied differences between relational grouping and grouping in the context of XML documents in [15]. A similar study about a cube operator is made by Wiwatwattana *et al.* in [17]. Borkar *et al.* [3] was one of the first to propose a precise syntax for an XQuery grouping clause. Beyer *et al.* [1] proposed a more powerful `group by` syntax for XQuery and presented use cases and performance analysis for data stored using a flat model. Kay [11] differentiated value grouping (OLAP) and positional grouping, based on nodes position within an XML document. He proposed a syntax for XQuery to support positional grouping. Lately, W3C published a working draft about the future version of XQuery 1.1 in [6] that proposes the addition of value grouping and windowing functionalities also studied in [4]. To the best of our knowledge, this paper is the first to compare the performance of grouping queries on different multidimensional models.

## 6   Conclusions and Future Work

We have analysed different approaches for representing multidimensional data in XML documents. We have proposed a hierarchical model which contains less redundancy and more semantics than the usual flat model and compared both to the XCube model from Hümmer *et al.* We have shown that queries on flat data models are always faster than those on other models because of the absence of complex joins. Similarly, we have shown that queries on the XCube model are the slowest because of the quantity of joins needed to traverse dimensions hierarchies.

Secondly, we have discussed how typical OLAP queries can be answered on data represented with those models. We have shown that common grouping operations are difficult to write and slow to execute in XQuery 1.0. However, by implementing the grouping clause proposed in XQuery 1.1 in the eXist XML native database, we have shown that these same queries could be made simpler to understand and faster to execute, making queries on the hierarchical model faster than those on the flat model without a grouping clause. We thus propose our hierarchical model associated with a grouping clause as a best compromise between redundancy and querying performance for data warehouses based on strict hierarchies.

Our experimentation was focussed on a straightforward implementation of the group-by clause with XQuery and the eXist open-source XML native database. It would be interesting to study how to optimise this implementation using more specialised index structures. It would also be interesting to compare eXist with other XML native database systems. Furthermore, another common type of data warehouse query is windowing: a windowing clause has been proposed in XQuery 1.1 and its implementation and evaluation are still open research problems. A longer term prospect is the creation of a general data warehouse benchmark for XML similar to the TCP-H relational data warehouse benchmark [16].